\documentclass[preprint,aps,prl,superscriptaddress]{revtex4}
\UseRawInputEncoding
\usepackage{multirow}
\usepackage{latexsym}
\usepackage{amssymb}
\usepackage{upgreek}
\usepackage{graphicx,epstopdf}
\usepackage{mathrsfs}
\usepackage{dcolumn}
\usepackage{bm,dsfont}
\usepackage{amsmath}
\usepackage{amsfonts}
\usepackage{color}
\usepackage{xcolor}
\usepackage{hyperref}
\hypersetup{
colorlinks = true,
linkcolor = [rgb]{0,0,1},
citecolor = [rgb]{0.75,0,0},
urlcolor = [rgb]{0.25, 0.21, 1}}

\newcommand{\eq}[1]{\begin{equation}#1\end{equation}}
\newcommand{\eqnref}[1]{Eq.\,\eqref{#1}}
\newcommand{\Eqref}[1]{Equation\,\eqref{#1}}
\newcommand{\figref}[1]{Fig.\,\ref{#1}}
\newcommand{\figsref}[1]{Figs.\,\ref{#1}}
\newcommand{\Figref}[1]{Figure\,\ref{#1}}

\begin{document}
\title{Restoration of non-Hermitian bulk-boundary correspondence by counterbalancing skin effect}
\author{Yi-Xin Xiao}
\affiliation{Department of Physics, Hong Kong University of Science and Technology, Clear Water Bay, Hong Kong, China}

\author{Zhao-Qing Zhang}
\affiliation{Department of Physics, Hong Kong University of Science and Technology, Clear Water Bay, Hong Kong, China}

\author{C. T. Chan}
\email{phchan@ust.hk}
\affiliation{Department of Physics, Hong Kong University of Science and Technology, Clear Water Bay, Hong Kong, China}


\begin{abstract}
The non-Hermitian skin effect (NHSE) undermines the conventional bulk-boundary correspondence (BBC) since it results in a distinct bulk spectrum in open-boundary systems compared to the periodic counterpart. Using the non-Hermitian (NH) Su-Schrieffer-Heeger (SSH) model as an example, we propose an intuitive approach, termed ``doubling and swapping" method, to restore the BBC. Explicitly, we construct a modified system by swapping the asymmetric intracell hoppings in every second primitive unit cell, such that it has double-sized unit cells compared to the NH SSH model and is free of NHSE. Importantly, the modified system and the NH SSH chain exhibit identical spectra under open boundary conditions (OBC). As a result, the modified system can serve as the valid bulk for defining topological invariants that correctly predicts edge states and topological phase transitions. The basic principle is applicable to many other systems such as the non-Hermitian Creutz ladder model. Furthermore, we extend the study to disordered systems in which the asymmetric hoppings are randomly swapped. We show that two types of winding numbers can be defined to account for the NHSE and topological edge states, respectively. 
\end{abstract}
\maketitle
{\color{blue} \it Introduction.---}Non-Hermitian (NH) Hamiltonians provide conceptually simple, intuitive, and powerful descriptions of a wide range of systems \cite{Ashida2020NHphysics}, such as wave systems with loss and gain \cite{ElGanainy2018review,Ozdemir2019review,feng2017review}, open systems \cite{Rotter2009openquantum,zhen2015EPring,cao2015Microcavities}. Diverse intriguing phenomena in non-Hermitian systems, such as exceptional points (EPs) \cite{miri2019EPreview}, have spurred extensive research and found numerous applications across various fields \cite{hodaei2014PTlasers,feng2014PTlaser,hodaei2017sensor,chen2017EPsensor}. 

The bulk-boundary correspondence (BBC) that predicts, for example, the chiral edge modes by the bulk Chern number is at the core of the topological band theory \cite{moore2021book}. The BBC implicitly assumes that the Hamiltonians with open boundary condition (OBC) share roughly the same bulk properties with their counterparts with periodic boundary condition (PBC). The assumption no longer holds in some non-Hermitian systems that exhibit non-Hermitian skin effect (NHSE) \cite{Lee2016Anomalous,Yao2018nonBloch,xiong2018Why,okuma2020Origin,Zhang2020winding}, which manifests as
the pileup of macroscopically many eigenstates at the system boundaries under OBC and the consequent contrast between OBC and PBC systems in their spectra. As a result, the conventional BBC needs to be rectified. 

Some remedies are proposed to resolve the problem, among which two approaches known as the non-Bloch approach \cite{Yao2018nonBloch,kunst2018Biorthogonal} and the biorthogonal approach \cite{kunst2018Biorthogonal} are most powerful and illuminating. The non-Bloch approach has been proved successful \cite{Yao2018nonBloch,yokomizo2019nonBloch,yang2020nonBloch}. By extending the concept of Brillouin zone (BZ) to generalized Brillouin zone (GBZ), a substitute Hamiltonian known as the non-Bloch Hamiltonian is constructed to serve as the bulk to define a topological invariant. The non-Bloch approach accurately predicts the existence of topological edge states and captures topological transitions. Its basic idea is to find a proper Hamiltonian that is devoid of NHSE and has identical OBC spectrum with the original system. The biorthogonal approach uses both the left and right eigenvectors to determine topological transitions and distinguish between edge states and bulk states, which exhibit a blurred distinction due to NHSE \cite{kunst2018Biorthogonal}. 

Here we propose a different approach which is physically intuitive, transparent, and notably less involved compared to the non-Bloch approach. For illustration, we consider the non-Hermitian (NH) Su-Schrieffer-Heeger (SSH) model with asymmetric intracell hoppings \cite{Yao2018nonBloch}. We adopt a double-sized unit cell comprising two primitive unit cells and swap the asymmetric intracell hoppings in every second primitive unit cell. This arrangement, which we call ``doubling and swapping", counterbalances the NHSE in the modified system. Importantly, we find that the ``doubling and swapping" does not change the OBC spectrum compared to the original NH SSH chain. Therefore the modified system can serve as the faithful bulk for defining topological invariants that correctly predicts edge states and topological phase transitions. The basic idea resembles that of the non-Bloch approach but is more straightforward and physically transparent: There is no need to figure out the GBZ and the only cost is involving more bands. We point out that the approach is also applicable to many other systems such as the PT-symmetric Creutz ladder with gain/loss \cite{Lee2016Anomalous}. Furthermore, we extend the study to disordered systems in which the asymmetric hoppings are randomly swapped. We find that two types of winding numbers can be defined in the real space to account for the presence of NHSE and topological edge states, respectively, in the disordered systems. 

\begin{figure}[htbp]
\includegraphics[clip, width=0.6\columnwidth, angle=0]{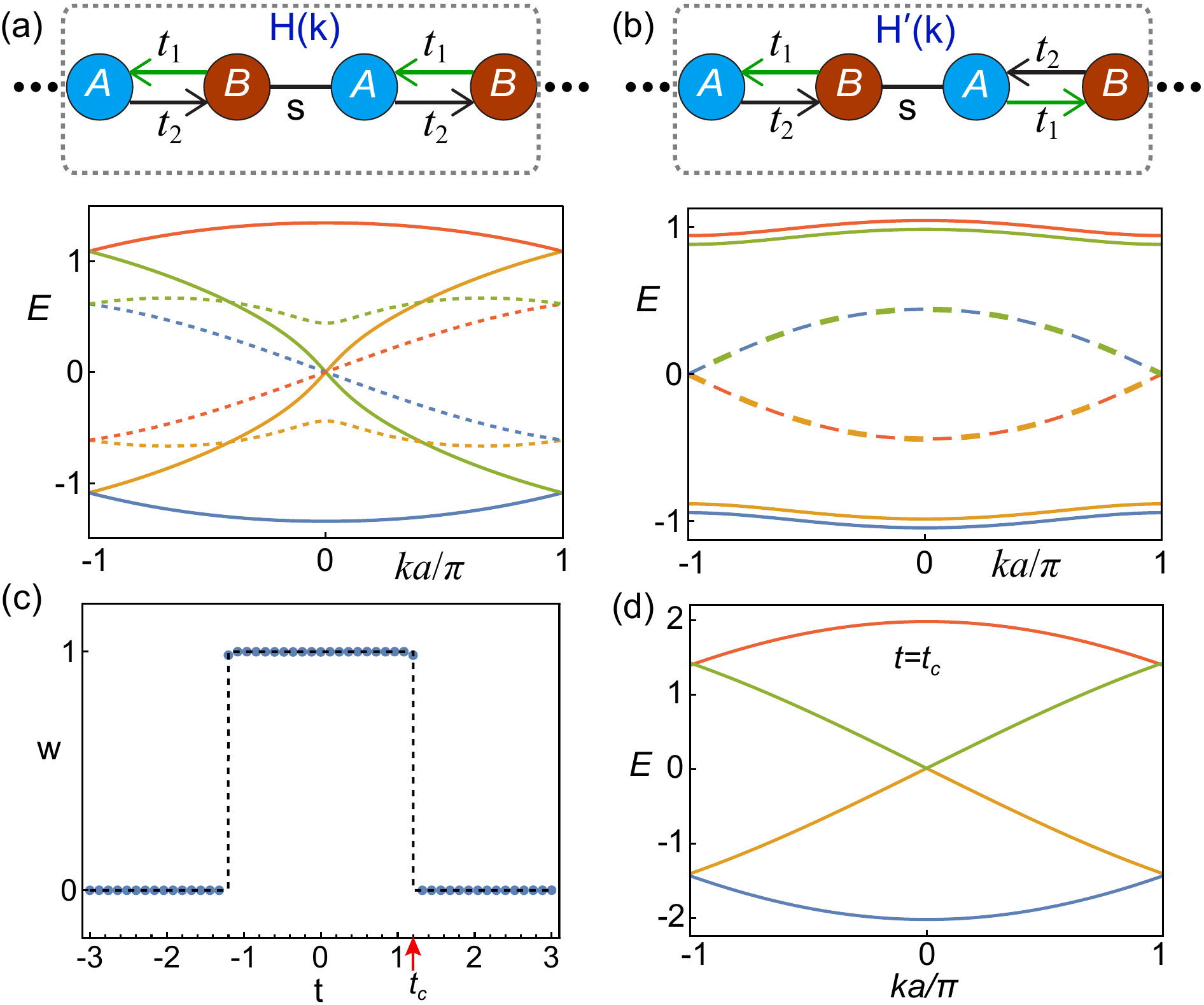}
\caption{(a) The supercell structure of the non-Hermitian SSH model and its band structure. $Re(E)$ and $Im(E)$ bands are represented by solid and dashed lines, respectively. 
(b) The modified system obtained from (a) by switching the non-reciprocal hoppings $t_1,\,t_2$ in every second unit cell, and its band structure. 
Four $Re(E)$ bands pairwise coincide, and for visual clarity they are intentionally drawn to be separated. 
$s=1,\, t_{1,2} = t \pm\gamma/2$ with $t=1/2,\, \gamma=4/3$ are used.  
(c) Variation of winding number $w$ as $t$ changes. 
(d) Bandgap closes at the topological transition point $t=t_c\approx 1.2$. 
\label{fig1}}
\end{figure}

{\color{blue} \it BBC restored by ``Doubling and swapping'' method.---}We consider the NH SSH model with asymmetric intracell hoppings  $t_1, t_2$ and intercell hopping $s$ shown in \figref{fig1}(a). The dashed box marks a double-sized unit cell consisting of two primitive unit cells, which corresponds to the four-band Bloch Hamiltonian 
\eq{
H(k)=\begin{pmatrix}
0&t_1&0&se^{-ika} \\ t_2&0&s&0 \\ 0&s&0&t_1 \\ se^{ika} &0&t_2&0
\end{pmatrix},
}
where $a=2 a_0$ with $a_0$ being the primitive lattice constant. It is well known that, due to NHSE, $H(k)$ is not the faithful bulk to capture the topological transitions as certain system parameter is tuned \cite{Yao2018nonBloch}. 

To counterbalance the NHSE, we swap the non-reciprocal hoppings $t_1, t_2$ in every second primitive cell and get a modified chain as shown in \figref{fig1}(b), which corresponds to a modified four-band Bloch Hamiltonian, 
\eq{
H'(k)=\begin{pmatrix}
0&t_1&0&se^{-ika} \\ t_2&0&s&0 \\ 0&s&0&t_2 \\ se^{ika}&0&t_1&0
\end{pmatrix}.
}
From a symmetry perspective, the presence of inversion symmetry in \figref{fig1} suppresses the NHSE.

For convenience, we refer to the OBC counterparts of $H(k)$ and $H'(k)$ as $H_{obc}$ and $H'_{obc}$, respectively. 
Just like Hermitian scenarios, $H'(k)$ can be used to establish the BBC to predict the topological edge states in $H'_{obc}$, due to the absence of NHSE. Being tridiagonal matrices that represent chains with nearest-neighbor hoppings, $H_{obc}$ and $H'_{obc}$ are equivalent under a similarity transformation and thus have identical spectrum \cite{Supp}. This implies that the NHSE-free system $H'(k)$ can serve as a valid bulk system for predicting edge states and topological transitions in $H_{obc}$ as well. 

The band structrues of $H(k)$ and $H'(k)$ are shown in \figsref{fig1}(a) and \ref{fig1}(b), where solid and dashed curves denote $ReE(k)$ and $ImE(k)$ bands, respectively, and $s=1,\, t_{1,2} = t \pm\gamma/2$ with $t=1/2,\, \gamma=4/3$ are assumed. The blue (red) solid band in \figref{fig1}(b) should actually coincide with the orange (green) band, and it is intentionally offset downward (upward) for better visibility. The bands in \figsref{fig1}(a) and \ref{fig1}(b) exhibit a clear contrast as expected. 

Due to chiral symmetry, $H'(k)$ can be flattened to $Q(k)=\mathbb{I}-2P(k)=\begin{pmatrix} 0 & q(k) \\ q^{-1}(k) & 0 \end{pmatrix}$, where $P(k)=\sum_{n=1,2} |\psi_n^R(k)\rangle \langle \psi_n^L(k)|$ with $\psi_n^{R/L}$ denoting right/left eigenvectors of $H'(k)$. The two-band winding number,
\eq{
w=\frac{1}{2\pi i} \int_{BZ} Tr \left[ q^{-1}(k)\partial_k q(k) \right] dk,
\label{w_k}
}
can be employed to capture the existence of topological edge states in $H_{obc}$.

\Figref{fig1}(c) depicts the topological transitions by jumps in the winding number $w$ at $t=\pm t_c=\pm \sqrt{s^2+(\gamma/2)^2}$ as $t$ varies, which agrees with the result of non-Bloch approach \cite{Yao2018nonBloch}. At the transition point $t=t_c$, $H^\prime(k)$ becomes gapless as is shown in \figref{fig1}(d). Note the bands become real-valued when $t_1 t_2\ge 0$, i.e., $|t| \ge \gamma/2$. 
When $t=\pm\gamma/2$, EPs occur at any $k\in(-\pi/a,\pi/a)$ since within each cell the coupling becomes unidirectional, while nondefective degeneracy points still remain at $k=\pm\pi/a$ \cite{Supp}. 
In addition to winding number $w$, we can use quantized Berry phase to characterize the topology and obtain a phase diagram similar to \figref{fig1}(c), utilizing the inversion symmetry in $H^\prime(k)$ \cite{Supp}.

Following a similar line of reasoning, we can also construct a two-band NHSE-free system,
\eq{
h^\prime (k)=\begin{pmatrix} 0& \tau+s e^{-ik a_0} \\ \tau+s e^{ik a_0} &0 \end{pmatrix}
\label{twoband}
} 
with $\tau=\sqrt{t_1 t_2}$, to serve as the bulk, since its OBC counterpart also shares the same spectrum with $H_{obc}$. We note that, $\tau=i\sqrt{|t_1 t_2|}$ and $h^\prime (k)$ is non-Hermitian when $t_1 t_2<0$. It is evident from \eqnref{twoband} that the nontrivial region $-t_c<t<t_c$ depicted in \figref{fig1}(c) can be inferred from $|\tau|<|s|$, meaning that the intracell hopping in $h^\prime (k)$ is weaker than the intercell hopping. We point out that $h^\prime (k)$ is just the non-Bloch Hamiltonian $H_{nB}(k)$ derived using the GBZ, i.e., $H_{nB}(k)=h^\prime (k)$ \cite{Yao2018nonBloch, xiao2022topology}. 

Compared to the non-Bloch approach, the ``doubling and swapping'' method is intuitive, transparent, and less involved, and meanwhile shares the advantage of enabling efficient computation of the continuous OBC spectrum of a large system, without the need for diagonalizing a large real-space Hamiltonian matrix which is especially challenging due to NHSE.
It is worth noting that the BBC in the non-Hermitian Creutz ladder model can also be restored similarly by swapping the gain and loss in every second primitive cell \cite{Lee2016Anomalous,Supp}, and the idea of counterbalancing NHSE to reestablish the BBC can be extended straightforwardly to higher dimensions \cite{Supp}. 

As for the presence of NHSE, a different winding number \cite{okuma2020Origin,Zhang2020winding} 
\eq{
\nu(E_b)=\frac{1}{2\pi i} \int_0^{2\pi} \frac{d}{dk} \ln \det \left[ H(k)-E_b \right] dk,
\label{nu_k}
}
can be used to characterize it, where $E_b$ denotes a reference point in the complex energy plane. 
To differentiate from \eqnref{w_k}, we call $\nu(E_b)$ the spectral winding number, and call $w$ as the chiral winding number.
Skin modes appear under OBC if and only if there is $E_b\in \mathbb{C}$ with respect to which the PBC spectrum $\sigma[H(k)]$ has nonzero winding, i.e., $\nu\ne 0$. In other words, a point gap under PBC is associated with the NHSE \cite{okuma2020Origin}.

\begin{figure}[htbp]
\includegraphics[clip, width=0.6\columnwidth, angle=0]{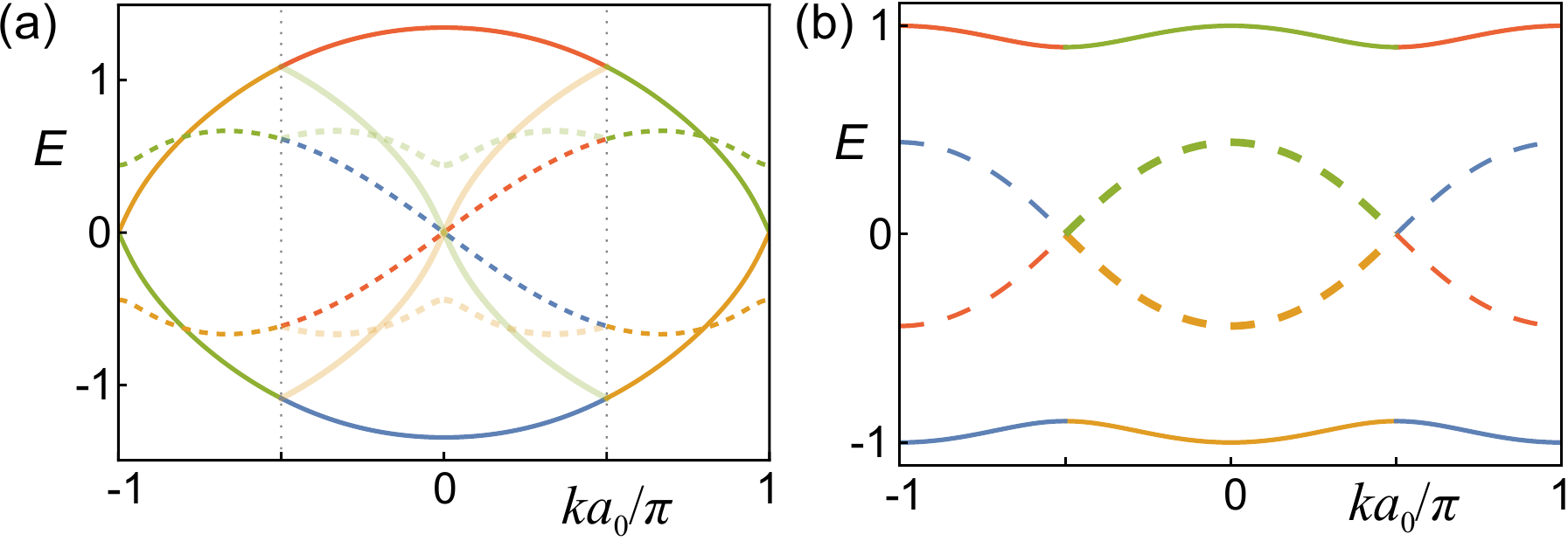}
\caption{(a) BZ folding of the $h(k)$ bands leads to the $H(k)$ bands in the $|ka_0/\pi|<1/2$ range, where translucent bands arise from BZ folding. Solid (dashed) curves denote $ReE$ ($ImE$) bands. 
(b) The $h^\prime(k)$ bands are derived from BZ unfolding of the $H^\prime(k)$ bands in \figref{fig1}(b), with its orange/blue bands and green/red bands becoming the 1st and 2nd band of $h^\prime(k)$. 
\label{fig2}}
\end{figure}

{\color{blue} \it Absence of BZ folding-induced degeneracy.---}We observe an unexpectd feature in the band structure shown in \figref{fig1}(a): the lack of degeneracy resulting from BZ folding, i.e., doubled unit cell. We use $h(k)$ to denote the two-band Bloch Hamiltonian associated with choosing the primitive unit cell in \figref{fig1}(a). The bands of $h(k)$ are shown in \figref{fig2}(a), where the translucent bands in the $|ka_0/\pi|<1/2$ range are shifted from $|ka_0/\pi|>1/2$ by BZ folding, and four colors are used to compare with \figref{fig1}(a). Obviously the folded bands in \figref{fig2}(a) coincide with that shown in \figref{fig1}(a). The NHSE causes $E_i(-k) \ne E_i(k)$ in \figref{fig1}(a), especially $E_i\left(-\frac{\pi}{a}\right) \ne E_i\left(\frac{\pi}{a}\right)$ noting $a=2a_0$, therefore the absence of BZ folding-induced degeneracy in \figref{fig2}(a). This is a unique property that cannot exist in NHSE-free systems. 

In contrast, the band structure of $H'(k)$ in \figref{fig1}(b) features degeneracies at the BZ boundaries, which can be obtained via BZ-folding the bands of the non-Bloch Hamiltonian $h_{nB}(k)\equiv h^\prime(k)$ shown in \figref{fig2}(b) \cite{Supp}. This implies that our intuitive method of swapping hoppings is equivalent to the non-Bloch approach.

\begin{figure}[htbp]
\includegraphics[clip, width=0.6\columnwidth, angle=0]{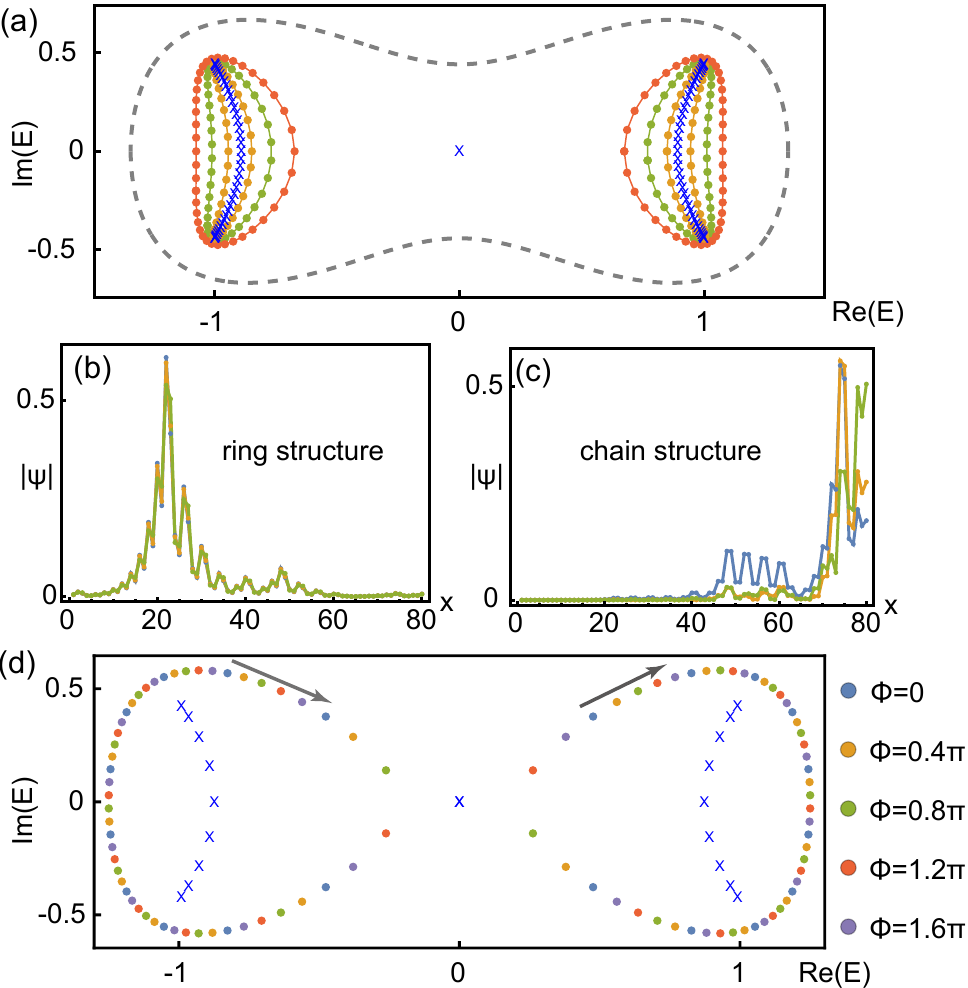}
\caption{(a) Spectra for the 1st, 2nd and 3rd $L=80$ random ring configurations $\mathcal{H}_r$ (orange, green, red), common spectrum for three chain configurations $\mathcal{H}_c$ (blue crossings), and $H(k)$'s spectrum (gray dashed). 
(b), (c) Three arbitrarily chosen wavefunctions $|\psi(x)|$ of the 3rd configuration for (b) $\mathcal{H}_r$ and (c) $\mathcal{H}_c$. 
(d) Depiction of winding number $\nu'(E)$ in \eqnref{nu_Phi} regarding NHSE: spectral winding induced by flux $\Phi$ increasing from $0$ to $2\pi$ in a $L=20$ random ring configuration. 
\label{fig3}}
\end{figure}

{\color{blue} \it Random swapping---}Next we perform random swaps of the asymmetric hoppings $t_1,\,t_2$ in an SSH chain, with an equal probability of $1/2$ for swapping or no action, and investigate how topology, NHSE and disorder interplay. We connect a disordered chain $\mathcal{H}_c$ end-to-end to get its ring counterpart $\mathcal{H}_r$. We denote the spectra of $\mathcal{H}_c$ and $\mathcal{H}_r$ by $\sigma[\mathcal{H}_r]$ and $\sigma[\mathcal{H}_c]$.

We observe that, akin to $\sigma[H(k)]$, $\sigma[\mathcal{H}_r]$ typically manifest as loops in the complex energy plane, which wind around the arcs of $\sigma[\mathcal{H}_c]$, as is shown in \figref{fig3}(a), where the spectra of the 1st, 2nd and 3rd $\mathcal{H}_r$ confirguations of size $L=80$ are denoted by orange, green and red dots, respectively. The three $\mathcal{H}_c$ confirguations share a common spectrum, including two edge states at $E\approx 0$,
as is denoted by the blue crossings in \figref{fig3}(a). The enclosing behavior in \figref{fig3}(a) implies remnant NHSE and the size of enclosed area signifies its strength. For comparison, $\sigma[H(k)]$ is shown by the large dashed circle, indicating much stronger NHSE under OBC than the disordered rings. We note $H'(k)$ is a special random configuration that is NHSE-free, and its spectrum $\sigma[H'(k)]$ encloses zero area. 

For illustration, three arbitrarily chosen wavefunctions of $\mathcal{H}_r$ and its chain counterpart $\mathcal{H}_c$ are shown in \figsref{fig3}(b) and \ref{fig3}(c), respectively, for the 3rd configuration which corresponds to the red dots in \figref{fig3}(a). 
With the ring cut open, wavefunctions accumulate to the right boundary as is shown in \figref{fig3}(c), indicating NHSE under swapping disorder. Different random chains are related by similarity transformations, as is the case for $H_{obc}$ and $H_{obc}'$. 

Similar to \eqnref{nu_k}, a winding number associated with the winding of $\sigma[\mathcal{H}_r]$ around $\sigma[\mathcal{H}_c]$ can be used to characetrize the NHSE \cite{gong2018PRX,claes2021SkinWinding}: 
\eq{
\nu'(E_b)=\frac{1}{{2\pi} i} \int_0^{2\pi} \frac{d}{d\Phi }\ln\det\left[ \mathcal{H}_r(\Phi)-E_b \right] d\Phi,
\label{nu_Phi}
}
where $\Phi$ denotes the flux that threads the $\mathcal{H}_r$ ring. \Eqref{nu_Phi} reduces to \eqnref{nu_k} when disorder is absent \cite{gong2018PRX}. The flux $\Phi$ appears as phase factors $e^{\pm i\Phi/L}$ that are multiplied to hoppings according to the Peierls substitution \cite{hofstadter1976magnetic}. Taking a $L=20$ random ring for example, the winding of $\sigma[\mathcal{H}_r(\Phi)]$ with respect to $\Phi$ is shown in \figref{fig3}(d), where five sets of spectra are displayed in different colors and the arrows indicate the directions of spectral winding. The random chain's spectrum $\sigma[\mathcal{H}_c]$ is shown by blue crossings, which is enclosed by $\sigma[\mathcal{H}_r(\Phi)]$.

\begin{figure}[htbp]
\includegraphics[clip, width=0.6\columnwidth, angle=0]{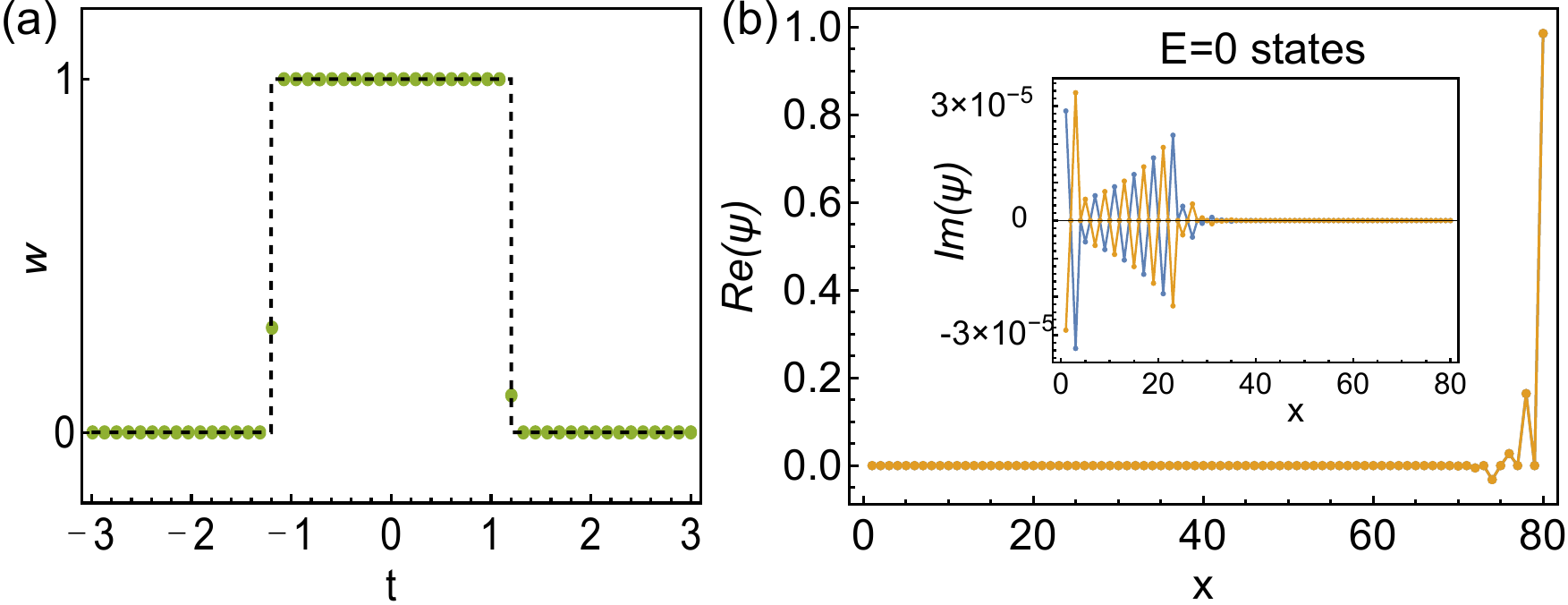}
\caption{(a) Winding number $w'$ of a $L=80$ random chain calculated using \eqnref{wreal}. 
(b) The two topologically protected $E=0$ states. 
\label{fig4}}
\end{figure}

Different random chains have identical spectrum and their topological edge states all occur around $E=0$ and are related by similarity transformations. Their common topological transitions and topological edge states can be captured by the winding number $w$ in \eqnref{w_k} defined based on $H'(k)$. Alternatively, the winding number for the random chains can also be computed using the real-space formula: \cite{song2019RealSpace,mondragonChiral2014}
\eq{
w'=\frac{1}{L^\prime} Tr^\prime (SQ[Q,X]),
\label{wreal}
}
where $S,\, Q$ and $X=diag\{1,1,2,2,\dots\}$ represent the chiral symmetry operator, flattened Hamiltonian and coordinate operator, respectively, and $Tr^\prime$ denotes the trace over the middle interval $x\in[\ell+1,L-\ell]$ of the chain. In practice, a modest chain size is sufficient. 
The variation of $w'$ with $t$ is shown in \figref{fig4}(a), which is the same as \figref{fig1}(c) except that the latter has the merit of being immune to finite-size errors. It can be demonstrated that the value of $w'$ remains unchanged when we alter the random configuration of $\mathcal{H}_c$ \cite{Supp}. This is expected because different random configurations of $\mathcal{H}_c$ yield the same spectrum.

$w'\ne 0$ predicts the existence of $E\approx 0$ states in a random chain, which are generally edge states and are related to the edge states of $H'_{obc}$ by a similarity transformation. For illustration, the two $E\approx 0$ edge states for the 3rd random $\mathcal{H}_c$ confirguation are shown in \figref{fig4}(b). 
They are both localized at the right edge with identical real parts, and differ slightly in their imaginary part, which are reminiscent of two states near an exceptional point. In contrast, each of the two $E\approx 0$ states of the $H'_{obc}$ chain is localized at both edges, exhibiting even/odd parity due to inversion symmetry, just like a Hermitian SSH chain. It turn out that the NHSE manifests in the similarity transformation, which amplifies (attenuates) the right (left) side of the even/odd-parity edge states' wavefunctions of $H'_{obc}$, and results in \figref{fig4}(b). 

{\color{blue} \it Conclusion.---}We introduced a simple and intuitive approach known as ``doubling and swapping" to restore the BBC in systems exhibiting NHSE, at the expense of involving more bands. The basic principle is to counterbalance the skin effect by swapping the asymmetric hoppings (or gain/loss) in every second primitive unit cell, meanwhile without changing the OBC spectrum. This approach is equivalent to the non-Bloch approach that uses the concept of GBZ, and may lead to the non-Bloch Hamiltonian directly. The idea can be applied to a wide range of systems, including those in higher dimensions. 
Furthermore, we extend the study to disordered systems with asymmetric hoppings randomly swapped. Like the ordered case, two types of winding numbers can be defined to account for the NHSE and topological edge states, respectively.

\begin{acknowledgments}
This work is supported by Research Grants Council (RGC) Hong Kong through grants 16303119 and 16307420.
\end{acknowledgments}


\begin{thebibliography}{27}%
\makeatletter
\providecommand \@ifxundefined [1]{%
 \@ifx{#1\undefined}
}%
\providecommand \@ifnum [1]{%
 \ifnum #1\expandafter \@firstoftwo
 \else \expandafter \@secondoftwo
 \fi
}%
\providecommand \@ifx [1]{%
 \ifx #1\expandafter \@firstoftwo
 \else \expandafter \@secondoftwo
 \fi
}%
\providecommand \natexlab [1]{#1}%
\providecommand \enquote  [1]{``#1''}%
\providecommand \bibnamefont  [1]{#1}%
\providecommand \bibfnamefont [1]{#1}%
\providecommand \citenamefont [1]{#1}%
\providecommand \href@noop [0]{\@secondoftwo}%
\providecommand \href [0]{\begingroup \@sanitize@url \@href}%
\providecommand \@href[1]{\@@startlink{#1}\@@href}%
\providecommand \@@href[1]{\endgroup#1\@@endlink}%
\providecommand \@sanitize@url [0]{\catcode `\\12\catcode `\$12\catcode
  `\&12\catcode `\#12\catcode `\^12\catcode `\_12\catcode `\%12\relax}%
\providecommand \@@startlink[1]{}%
\providecommand \@@endlink[0]{}%
\providecommand \url  [0]{\begingroup\@sanitize@url \@url }%
\providecommand \@url [1]{\endgroup\@href {#1}{\urlprefix }}%
\providecommand \urlprefix  [0]{URL }%
\providecommand \Eprint [0]{\href }%
\providecommand \doibase [0]{https://doi.org/}%
\providecommand \selectlanguage [0]{\@gobble}%
\providecommand \bibinfo  [0]{\@secondoftwo}%
\providecommand \bibfield  [0]{\@secondoftwo}%
\providecommand \translation [1]{[#1]}%
\providecommand \BibitemOpen [0]{}%
\providecommand \bibitemStop [0]{}%
\providecommand \bibitemNoStop [0]{.\EOS\space}%
\providecommand \EOS [0]{\spacefactor3000\relax}%
\providecommand \BibitemShut  [1]{\csname bibitem#1\endcsname}%
\let\auto@bib@innerbib\@empty
\bibitem [{\citenamefont {Ashida}\ \emph {et~al.}(2020)\citenamefont {Ashida},
  \citenamefont {Gong},\ and\ \citenamefont {Ueda}}]{Ashida2020NHphysics}%
  \BibitemOpen
  \bibfield  {author} {\bibinfo {author} {\bibfnamefont {Y.}~\bibnamefont
  {Ashida}}, \bibinfo {author} {\bibfnamefont {Z.}~\bibnamefont {Gong}},\ and\
  \bibinfo {author} {\bibfnamefont {M.}~\bibnamefont {Ueda}},\ }\bibfield
  {title} {\bibinfo {title} {Non-{{Hermitian}} physics},\ }\href
  {https://doi.org/10.1080/00018732.2021.1876991} {\bibfield  {journal}
  {\bibinfo  {journal} {Adv. Phys.}\ }\textbf {\bibinfo {volume} {69}},\
  \bibinfo {pages} {249} (\bibinfo {year} {2020})}\BibitemShut {NoStop}%
\bibitem [{\citenamefont {{El-Ganainy}}\ \emph {et~al.}(2018)\citenamefont
  {{El-Ganainy}}, \citenamefont {Makris}, \citenamefont {Khajavikhan},
  \citenamefont {Musslimani}, \citenamefont {Rotter},\ and\ \citenamefont
  {Christodoulides}}]{ElGanainy2018review}%
  \BibitemOpen
  \bibfield  {author} {\bibinfo {author} {\bibfnamefont {R.}~\bibnamefont
  {{El-Ganainy}}}, \bibinfo {author} {\bibfnamefont {K.~G.}\ \bibnamefont
  {Makris}}, \bibinfo {author} {\bibfnamefont {M.}~\bibnamefont {Khajavikhan}},
  \bibinfo {author} {\bibfnamefont {Z.~H.}\ \bibnamefont {Musslimani}},
  \bibinfo {author} {\bibfnamefont {S.}~\bibnamefont {Rotter}},\ and\ \bibinfo
  {author} {\bibfnamefont {D.~N.}\ \bibnamefont {Christodoulides}},\ }\bibfield
   {title} {\bibinfo {title} {Non-{{Hermitian}} physics and {{PT}} symmetry},\
  }\href {https://doi.org/10.1038/nphys4323} {\bibfield  {journal} {\bibinfo
  {journal} {Nature Physics}\ }\textbf {\bibinfo {volume} {14}},\ \bibinfo
  {pages} {11} (\bibinfo {year} {2018})}\BibitemShut {NoStop}%
\bibitem [{\citenamefont {{\"O}zdemir}\ \emph {et~al.}(2019)\citenamefont
  {{\"O}zdemir}, \citenamefont {Rotter}, \citenamefont {Nori},\ and\
  \citenamefont {Yang}}]{Ozdemir2019review}%
  \BibitemOpen
  \bibfield  {author} {\bibinfo {author} {\bibfnamefont {{\c S}.~K.}\
  \bibnamefont {{\"O}zdemir}}, \bibinfo {author} {\bibfnamefont
  {S.}~\bibnamefont {Rotter}}, \bibinfo {author} {\bibfnamefont
  {F.}~\bibnamefont {Nori}},\ and\ \bibinfo {author} {\bibfnamefont
  {L.}~\bibnamefont {Yang}},\ }\bibfield  {title} {\bibinfo {title}
  {Parity\textendash time symmetry and exceptional points in photonics},\
  }\href {https://doi.org/10.1038/s41563-019-0304-9} {\bibfield  {journal}
  {\bibinfo  {journal} {Nat. Mater.}\ }\textbf {\bibinfo {volume} {18}},\
  \bibinfo {pages} {783} (\bibinfo {year} {2019})}\BibitemShut {NoStop}%
\bibitem [{\citenamefont {Feng}\ \emph {et~al.}(2017)\citenamefont {Feng},
  \citenamefont {{El-Ganainy}},\ and\ \citenamefont {Ge}}]{feng2017review}%
  \BibitemOpen
  \bibfield  {author} {\bibinfo {author} {\bibfnamefont {L.}~\bibnamefont
  {Feng}}, \bibinfo {author} {\bibfnamefont {R.}~\bibnamefont {{El-Ganainy}}},\
  and\ \bibinfo {author} {\bibfnamefont {L.}~\bibnamefont {Ge}},\ }\bibfield
  {title} {\bibinfo {title} {Non-{{Hermitian}} photonics based on
  parity\textendash time symmetry},\ }\href
  {https://doi.org/10.1038/s41566-017-0031-1} {\bibfield  {journal} {\bibinfo
  {journal} {Nature Photon}\ }\textbf {\bibinfo {volume} {11}},\ \bibinfo
  {pages} {752} (\bibinfo {year} {2017})}\BibitemShut {NoStop}%
\bibitem [{\citenamefont {Rotter}(2009)}]{Rotter2009openquantum}%
  \BibitemOpen
  \bibfield  {author} {\bibinfo {author} {\bibfnamefont {I.}~\bibnamefont
  {Rotter}},\ }\bibfield  {title} {\bibinfo {title} {A non-{{Hermitian
  Hamilton}} operator and the physics of open quantum systems},\ }\href
  {https://doi.org/10.1088/1751-8113/42/15/153001} {\bibfield  {journal}
  {\bibinfo  {journal} {J. Phys. A: Math. Theor.}\ }\textbf {\bibinfo {volume}
  {42}},\ \bibinfo {pages} {153001} (\bibinfo {year} {2009})}\BibitemShut
  {NoStop}%
\bibitem [{\citenamefont {Zhen}\ \emph {et~al.}(2015)\citenamefont {Zhen},
  \citenamefont {Hsu}, \citenamefont {Igarashi}, \citenamefont {Lu},
  \citenamefont {Kaminer}, \citenamefont {Pick}, \citenamefont {Chua},
  \citenamefont {Joannopoulos},\ and\ \citenamefont {Solja{\v
  c}i{\'c}}}]{zhen2015EPring}%
  \BibitemOpen
  \bibfield  {author} {\bibinfo {author} {\bibfnamefont {B.}~\bibnamefont
  {Zhen}}, \bibinfo {author} {\bibfnamefont {C.~W.}\ \bibnamefont {Hsu}},
  \bibinfo {author} {\bibfnamefont {Y.}~\bibnamefont {Igarashi}}, \bibinfo
  {author} {\bibfnamefont {L.}~\bibnamefont {Lu}}, \bibinfo {author}
  {\bibfnamefont {I.}~\bibnamefont {Kaminer}}, \bibinfo {author} {\bibfnamefont
  {A.}~\bibnamefont {Pick}}, \bibinfo {author} {\bibfnamefont {S.-L.}\
  \bibnamefont {Chua}}, \bibinfo {author} {\bibfnamefont {J.~D.}\ \bibnamefont
  {Joannopoulos}},\ and\ \bibinfo {author} {\bibfnamefont {M.}~\bibnamefont
  {Solja{\v c}i{\'c}}},\ }\bibfield  {title} {\bibinfo {title} {Spawning rings
  of exceptional points out of {{Dirac}} cones},\ }\href
  {https://doi.org/10.1038/nature14889} {\bibfield  {journal} {\bibinfo
  {journal} {Nature (London)}\ }\textbf {\bibinfo {volume} {525}},\ \bibinfo
  {pages} {354} (\bibinfo {year} {2015})}\BibitemShut {NoStop}%
\bibitem [{\citenamefont {Cao}\ and\ \citenamefont
  {Wiersig}(2015)}]{cao2015Microcavities}%
  \BibitemOpen
  \bibfield  {author} {\bibinfo {author} {\bibfnamefont {H.}~\bibnamefont
  {Cao}}\ and\ \bibinfo {author} {\bibfnamefont {J.}~\bibnamefont {Wiersig}},\
  }\bibfield  {title} {\bibinfo {title} {Dielectric microcavities: {{Model}}
  systems for wave chaos and non-{{Hermitian}} physics},\ }\href
  {https://doi.org/10.1103/RevModPhys.87.61} {\bibfield  {journal} {\bibinfo
  {journal} {Rev. Mod. Phys.}\ }\textbf {\bibinfo {volume} {87}},\ \bibinfo
  {pages} {61} (\bibinfo {year} {2015})}\BibitemShut {NoStop}%
\bibitem [{\citenamefont {Miri}\ and\ \citenamefont
  {Al{\`u}}(2019)}]{miri2019EPreview}%
  \BibitemOpen
  \bibfield  {author} {\bibinfo {author} {\bibfnamefont {M.-A.}\ \bibnamefont
  {Miri}}\ and\ \bibinfo {author} {\bibfnamefont {A.}~\bibnamefont {Al{\`u}}},\
  }\bibfield  {title} {\bibinfo {title} {Exceptional points in optics and
  photonics},\ }\href {https://doi.org/10.1126/science.aar7709} {\bibfield
  {journal} {\bibinfo  {journal} {Science}\ }\textbf {\bibinfo {volume}
  {363}},\ \bibinfo {pages} {eaar7709} (\bibinfo {year} {2019})}\BibitemShut
  {NoStop}%
\bibitem [{\citenamefont {Hodaei}\ \emph {et~al.}(2014)\citenamefont {Hodaei},
  \citenamefont {Miri}, \citenamefont {Heinrich}, \citenamefont
  {Christodoulides},\ and\ \citenamefont {Khajavikhan}}]{hodaei2014PTlasers}%
  \BibitemOpen
  \bibfield  {author} {\bibinfo {author} {\bibfnamefont {H.}~\bibnamefont
  {Hodaei}}, \bibinfo {author} {\bibfnamefont {M.-A.}\ \bibnamefont {Miri}},
  \bibinfo {author} {\bibfnamefont {M.}~\bibnamefont {Heinrich}}, \bibinfo
  {author} {\bibfnamefont {D.~N.}\ \bibnamefont {Christodoulides}},\ and\
  \bibinfo {author} {\bibfnamefont {M.}~\bibnamefont {Khajavikhan}},\
  }\bibfield  {title} {\bibinfo {title} {Parity-time\textendash symmetric
  microring lasers},\ }\href {https://doi.org/10.1126/science.1258480}
  {\bibfield  {journal} {\bibinfo  {journal} {Science}\ }\textbf {\bibinfo
  {volume} {346}},\ \bibinfo {pages} {975} (\bibinfo {year}
  {2014})}\BibitemShut {NoStop}%
\bibitem [{\citenamefont {Feng}\ \emph {et~al.}(2014)\citenamefont {Feng},
  \citenamefont {Wong}, \citenamefont {Ma}, \citenamefont {Wang},\ and\
  \citenamefont {Zhang}}]{feng2014PTlaser}%
  \BibitemOpen
  \bibfield  {author} {\bibinfo {author} {\bibfnamefont {L.}~\bibnamefont
  {Feng}}, \bibinfo {author} {\bibfnamefont {Z.~J.}\ \bibnamefont {Wong}},
  \bibinfo {author} {\bibfnamefont {R.-M.}\ \bibnamefont {Ma}}, \bibinfo
  {author} {\bibfnamefont {Y.}~\bibnamefont {Wang}},\ and\ \bibinfo {author}
  {\bibfnamefont {X.}~\bibnamefont {Zhang}},\ }\bibfield  {title} {\bibinfo
  {title} {Single-mode laser by parity-time symmetry breaking},\ }\href
  {https://doi.org/10.1126/science.1258479} {\bibfield  {journal} {\bibinfo
  {journal} {Science}\ }\textbf {\bibinfo {volume} {346}},\ \bibinfo {pages}
  {972} (\bibinfo {year} {2014})}\BibitemShut {NoStop}%
\bibitem [{\citenamefont {Hodaei}\ \emph {et~al.}(2017)\citenamefont {Hodaei},
  \citenamefont {Hassan}, \citenamefont {Wittek}, \citenamefont
  {{Garcia-Gracia}}, \citenamefont {{El-Ganainy}}, \citenamefont
  {Christodoulides},\ and\ \citenamefont {Khajavikhan}}]{hodaei2017sensor}%
  \BibitemOpen
  \bibfield  {author} {\bibinfo {author} {\bibfnamefont {H.}~\bibnamefont
  {Hodaei}}, \bibinfo {author} {\bibfnamefont {A.~U.}\ \bibnamefont {Hassan}},
  \bibinfo {author} {\bibfnamefont {S.}~\bibnamefont {Wittek}}, \bibinfo
  {author} {\bibfnamefont {H.}~\bibnamefont {{Garcia-Gracia}}}, \bibinfo
  {author} {\bibfnamefont {R.}~\bibnamefont {{El-Ganainy}}}, \bibinfo {author}
  {\bibfnamefont {D.~N.}\ \bibnamefont {Christodoulides}},\ and\ \bibinfo
  {author} {\bibfnamefont {M.}~\bibnamefont {Khajavikhan}},\ }\bibfield
  {title} {\bibinfo {title} {Enhanced sensitivity at higher-order exceptional
  points},\ }\href {https://doi.org/10.1038/nature23280} {\bibfield  {journal}
  {\bibinfo  {journal} {Nature (London)}\ }\textbf {\bibinfo {volume} {548}},\
  \bibinfo {pages} {187} (\bibinfo {year} {2017})}\BibitemShut {NoStop}%
\bibitem [{\citenamefont {Chen}\ \emph {et~al.}(2017)\citenamefont {Chen},
  \citenamefont {{\"O}zdemir}, \citenamefont {Zhao}, \citenamefont {Wiersig},\
  and\ \citenamefont {Yang}}]{chen2017EPsensor}%
  \BibitemOpen
  \bibfield  {author} {\bibinfo {author} {\bibfnamefont {W.}~\bibnamefont
  {Chen}}, \bibinfo {author} {\bibfnamefont {{\c S}.~K.}\ \bibnamefont
  {{\"O}zdemir}}, \bibinfo {author} {\bibfnamefont {G.}~\bibnamefont {Zhao}},
  \bibinfo {author} {\bibfnamefont {J.}~\bibnamefont {Wiersig}},\ and\ \bibinfo
  {author} {\bibfnamefont {L.}~\bibnamefont {Yang}},\ }\bibfield  {title}
  {\bibinfo {title} {Exceptional points enhance sensing in an optical
  microcavity},\ }\href {https://doi.org/10.1038/nature23281} {\bibfield
  {journal} {\bibinfo  {journal} {Nature}\ }\textbf {\bibinfo {volume} {548}},\
  \bibinfo {pages} {192} (\bibinfo {year} {2017})}\BibitemShut {NoStop}%
\bibitem [{\citenamefont {Moessner}\ and\ \citenamefont
  {Moore}(2021)}]{moore2021book}%
  \BibitemOpen
  \bibfield  {author} {\bibinfo {author} {\bibfnamefont {R.}~\bibnamefont
  {Moessner}}\ and\ \bibinfo {author} {\bibfnamefont {J.~E.}\ \bibnamefont
  {Moore}},\ }\href@noop {} {\emph {\bibinfo {title} {Topological {{Phases}} of
  {{Matter}}}}}\ (\bibinfo  {publisher} {{Cambridge University Press}},\
  \bibinfo {address} {{Cambridge}},\ \bibinfo {year} {2021})\BibitemShut
  {NoStop}%
\bibitem [{\citenamefont {Lee}(2016)}]{Lee2016Anomalous}%
  \BibitemOpen
  \bibfield  {author} {\bibinfo {author} {\bibfnamefont {T.~E.}\ \bibnamefont
  {Lee}},\ }\bibfield  {title} {\bibinfo {title} {Anomalous {{Edge State}} in a
  {{Non-Hermitian Lattice}}},\ }\href
  {https://doi.org/10.1103/PhysRevLett.116.133903} {\bibfield  {journal}
  {\bibinfo  {journal} {Phys. Rev. Lett.}\ }\textbf {\bibinfo {volume} {116}},\
  \bibinfo {pages} {133903} (\bibinfo {year} {2016})}\BibitemShut {NoStop}%
\bibitem [{\citenamefont {Yao}\ and\ \citenamefont
  {Wang}(2018)}]{Yao2018nonBloch}%
  \BibitemOpen
  \bibfield  {author} {\bibinfo {author} {\bibfnamefont {S.}~\bibnamefont
  {Yao}}\ and\ \bibinfo {author} {\bibfnamefont {Z.}~\bibnamefont {Wang}},\
  }\bibfield  {title} {\bibinfo {title} {Edge {{States}} and {{Topological
  Invariants}} of {{Non-Hermitian Systems}}},\ }\href
  {https://doi.org/10.1103/PhysRevLett.121.086803} {\bibfield  {journal}
  {\bibinfo  {journal} {Phys. Rev. Lett.}\ }\textbf {\bibinfo {volume} {121}},\
  \bibinfo {pages} {086803} (\bibinfo {year} {2018})}\BibitemShut {NoStop}%
\bibitem [{\citenamefont {Xiong}(2018)}]{xiong2018Why}%
  \BibitemOpen
  \bibfield  {author} {\bibinfo {author} {\bibfnamefont {Y.}~\bibnamefont
  {Xiong}},\ }\bibfield  {title} {\bibinfo {title} {Why does bulk boundary
  correspondence fail in some non-hermitian topological models},\ }\href
  {https://doi.org/10.1088/2399-6528/aab64a} {\bibfield  {journal} {\bibinfo
  {journal} {J. Phys. Commun.}\ }\textbf {\bibinfo {volume} {2}},\ \bibinfo
  {pages} {035043} (\bibinfo {year} {2018})}\BibitemShut {NoStop}%
\bibitem [{\citenamefont {Okuma}\ \emph {et~al.}(2020)\citenamefont {Okuma},
  \citenamefont {Kawabata}, \citenamefont {Shiozaki},\ and\ \citenamefont
  {Sato}}]{okuma2020Origin}%
  \BibitemOpen
  \bibfield  {author} {\bibinfo {author} {\bibfnamefont {N.}~\bibnamefont
  {Okuma}}, \bibinfo {author} {\bibfnamefont {K.}~\bibnamefont {Kawabata}},
  \bibinfo {author} {\bibfnamefont {K.}~\bibnamefont {Shiozaki}},\ and\
  \bibinfo {author} {\bibfnamefont {M.}~\bibnamefont {Sato}},\ }\bibfield
  {title} {\bibinfo {title} {Topological {{Origin}} of {{Non-Hermitian Skin
  Effects}}},\ }\href {https://doi.org/10.1103/PhysRevLett.124.086801}
  {\bibfield  {journal} {\bibinfo  {journal} {Phys. Rev. Lett.}\ }\textbf
  {\bibinfo {volume} {124}},\ \bibinfo {pages} {086801} (\bibinfo {year}
  {2020})}\BibitemShut {NoStop}%
\bibitem [{\citenamefont {Zhang}\ \emph {et~al.}(2020)\citenamefont {Zhang},
  \citenamefont {Yang},\ and\ \citenamefont {Fang}}]{Zhang2020winding}%
  \BibitemOpen
  \bibfield  {author} {\bibinfo {author} {\bibfnamefont {K.}~\bibnamefont
  {Zhang}}, \bibinfo {author} {\bibfnamefont {Z.}~\bibnamefont {Yang}},\ and\
  \bibinfo {author} {\bibfnamefont {C.}~\bibnamefont {Fang}},\ }\bibfield
  {title} {\bibinfo {title} {Correspondence between {{Winding Numbers}} and
  {{Skin Modes}} in {{Non-Hermitian Systems}}},\ }\href
  {https://doi.org/10.1103/PhysRevLett.125.126402} {\bibfield  {journal}
  {\bibinfo  {journal} {Phys. Rev. Lett.}\ }\textbf {\bibinfo {volume} {125}},\
  \bibinfo {pages} {126402} (\bibinfo {year} {2020})}\BibitemShut {NoStop}%
\bibitem [{\citenamefont {Kunst}\ \emph {et~al.}(2018)\citenamefont {Kunst},
  \citenamefont {Edvardsson}, \citenamefont {Budich},\ and\ \citenamefont
  {Bergholtz}}]{kunst2018Biorthogonal}%
  \BibitemOpen
  \bibfield  {author} {\bibinfo {author} {\bibfnamefont {F.~K.}\ \bibnamefont
  {Kunst}}, \bibinfo {author} {\bibfnamefont {E.}~\bibnamefont {Edvardsson}},
  \bibinfo {author} {\bibfnamefont {J.~C.}\ \bibnamefont {Budich}},\ and\
  \bibinfo {author} {\bibfnamefont {E.~J.}\ \bibnamefont {Bergholtz}},\
  }\bibfield  {title} {\bibinfo {title} {Biorthogonal {{Bulk-Boundary
  Correspondence}} in {{Non-Hermitian Systems}}},\ }\href
  {https://doi.org/10.1103/PhysRevLett.121.026808} {\bibfield  {journal}
  {\bibinfo  {journal} {Phys. Rev. Lett.}\ }\textbf {\bibinfo {volume} {121}},\
  \bibinfo {pages} {026808} (\bibinfo {year} {2018})}\BibitemShut {NoStop}%
\bibitem [{\citenamefont {Yokomizo}\ and\ \citenamefont
  {Murakami}(2019)}]{yokomizo2019nonBloch}%
  \BibitemOpen
  \bibfield  {author} {\bibinfo {author} {\bibfnamefont {K.}~\bibnamefont
  {Yokomizo}}\ and\ \bibinfo {author} {\bibfnamefont {S.}~\bibnamefont
  {Murakami}},\ }\bibfield  {title} {\bibinfo {title} {Non-{{Bloch Band
  Theory}} of {{Non-Hermitian Systems}}},\ }\href
  {https://doi.org/10.1103/PhysRevLett.123.066404} {\bibfield  {journal}
  {\bibinfo  {journal} {Phys. Rev. Lett.}\ }\textbf {\bibinfo {volume} {123}},\
  \bibinfo {pages} {066404} (\bibinfo {year} {2019})}\BibitemShut {NoStop}%
\bibitem [{\citenamefont {Yang}\ \emph {et~al.}(2020)\citenamefont {Yang},
  \citenamefont {Zhang}, \citenamefont {Fang},\ and\ \citenamefont
  {Hu}}]{yang2020nonBloch}%
  \BibitemOpen
  \bibfield  {author} {\bibinfo {author} {\bibfnamefont {Z.}~\bibnamefont
  {Yang}}, \bibinfo {author} {\bibfnamefont {K.}~\bibnamefont {Zhang}},
  \bibinfo {author} {\bibfnamefont {C.}~\bibnamefont {Fang}},\ and\ \bibinfo
  {author} {\bibfnamefont {J.}~\bibnamefont {Hu}},\ }\bibfield  {title}
  {\bibinfo {title} {Non-{{Hermitian Bulk-Boundary Correspondence}} and
  {{Auxiliary Generalized Brillouin Zone Theory}}},\ }\href
  {https://doi.org/10.1103/PhysRevLett.125.226402} {\bibfield  {journal}
  {\bibinfo  {journal} {Phys. Rev. Lett.}\ }\textbf {\bibinfo {volume} {125}},\
  \bibinfo {pages} {226402} (\bibinfo {year} {2020})}\BibitemShut {NoStop}%
\bibitem [{\citenamefont {Xiao}\ and\ \citenamefont
  {Chan}(2022)}]{xiao2022topology}%
  \BibitemOpen
  \bibfield  {author} {\bibinfo {author} {\bibfnamefont {Y.-X.}\ \bibnamefont
  {Xiao}}\ and\ \bibinfo {author} {\bibfnamefont {C.~T.}\ \bibnamefont
  {Chan}},\ }\bibfield  {title} {\bibinfo {title} {Topology in non-{{Hermitian
  Chern}} insulators with skin effect},\ }\href
  {https://doi.org/10.1103/PhysRevB.105.075128} {\bibfield  {journal} {\bibinfo
   {journal} {Phys. Rev. B}\ }\textbf {\bibinfo {volume} {105}},\ \bibinfo
  {pages} {075128} (\bibinfo {year} {2022})}\BibitemShut {NoStop}%
\bibitem [{\citenamefont {Gong}\ \emph {et~al.}(2018)\citenamefont {Gong},
  \citenamefont {Ashida}, \citenamefont {Kawabata}, \citenamefont {Takasan},
  \citenamefont {Higashikawa},\ and\ \citenamefont {Ueda}}]{gong2018PRX}%
  \BibitemOpen
  \bibfield  {author} {\bibinfo {author} {\bibfnamefont {Z.}~\bibnamefont
  {Gong}}, \bibinfo {author} {\bibfnamefont {Y.}~\bibnamefont {Ashida}},
  \bibinfo {author} {\bibfnamefont {K.}~\bibnamefont {Kawabata}}, \bibinfo
  {author} {\bibfnamefont {K.}~\bibnamefont {Takasan}}, \bibinfo {author}
  {\bibfnamefont {S.}~\bibnamefont {Higashikawa}},\ and\ \bibinfo {author}
  {\bibfnamefont {M.}~\bibnamefont {Ueda}},\ }\bibfield  {title} {\bibinfo
  {title} {Topological {{Phases}} of {{Non-Hermitian Systems}}},\ }\href
  {https://doi.org/10.1103/PhysRevX.8.031079} {\bibfield  {journal} {\bibinfo
  {journal} {Phys. Rev. X}\ }\textbf {\bibinfo {volume} {8}},\ \bibinfo {pages}
  {031079} (\bibinfo {year} {2018})}\BibitemShut {NoStop}%
\bibitem [{\citenamefont {Claes}\ and\ \citenamefont
  {Hughes}(2021)}]{claes2021SkinWinding}%
  \BibitemOpen
  \bibfield  {author} {\bibinfo {author} {\bibfnamefont {J.}~\bibnamefont
  {Claes}}\ and\ \bibinfo {author} {\bibfnamefont {T.~L.}\ \bibnamefont
  {Hughes}},\ }\bibfield  {title} {\bibinfo {title} {Skin effect and winding
  number in disordered non-{{Hermitian}} systems},\ }\href
  {https://doi.org/10.1103/PhysRevB.103.L140201} {\bibfield  {journal}
  {\bibinfo  {journal} {Phys. Rev. B}\ }\textbf {\bibinfo {volume} {103}},\
  \bibinfo {pages} {L140201} (\bibinfo {year} {2021})}\BibitemShut {NoStop}%
\bibitem [{\citenamefont {Hofstadter}(1976)}]{hofstadter1976magnetic}%
  \BibitemOpen
  \bibfield  {author} {\bibinfo {author} {\bibfnamefont {D.~R.}\ \bibnamefont
  {Hofstadter}},\ }\bibfield  {title} {\bibinfo {title} {Energy levels and wave
  functions of {{Bloch}} electrons in rational and irrational magnetic
  fields},\ }\href {https://doi.org/10.1103/PhysRevB.14.2239} {\bibfield
  {journal} {\bibinfo  {journal} {Phys. Rev. B}\ }\textbf {\bibinfo {volume}
  {14}},\ \bibinfo {pages} {2239} (\bibinfo {year} {1976})}\BibitemShut
  {NoStop}%
\bibitem [{\citenamefont {Song}\ \emph {et~al.}(2019)\citenamefont {Song},
  \citenamefont {Yao},\ and\ \citenamefont {Wang}}]{song2019RealSpace}%
  \BibitemOpen
  \bibfield  {author} {\bibinfo {author} {\bibfnamefont {F.}~\bibnamefont
  {Song}}, \bibinfo {author} {\bibfnamefont {S.}~\bibnamefont {Yao}},\ and\
  \bibinfo {author} {\bibfnamefont {Z.}~\bibnamefont {Wang}},\ }\bibfield
  {title} {\bibinfo {title} {Non-{{Hermitian Topological Invariants}} in {{Real
  Space}}},\ }\href {https://doi.org/10.1103/PhysRevLett.123.246801} {\bibfield
   {journal} {\bibinfo  {journal} {Phys. Rev. Lett.}\ }\textbf {\bibinfo
  {volume} {123}},\ \bibinfo {pages} {246801} (\bibinfo {year}
  {2019})}\BibitemShut {NoStop}%
\bibitem [{\citenamefont {{Mondragon-Shem}}\ \emph {et~al.}(2014)\citenamefont
  {{Mondragon-Shem}}, \citenamefont {Hughes}, \citenamefont {Song},\ and\
  \citenamefont {Prodan}}]{mondragonChiral2014}%
  \BibitemOpen
  \bibfield  {author} {\bibinfo {author} {\bibfnamefont {I.}~\bibnamefont
  {{Mondragon-Shem}}}, \bibinfo {author} {\bibfnamefont {T.~L.}\ \bibnamefont
  {Hughes}}, \bibinfo {author} {\bibfnamefont {J.}~\bibnamefont {Song}},\ and\
  \bibinfo {author} {\bibfnamefont {E.}~\bibnamefont {Prodan}},\ }\bibfield
  {title} {\bibinfo {title} {Topological {{Criticality}} in the
  {{Chiral-Symmetric AIII Class}} at {{Strong Disorder}}},\ }\href
  {https://doi.org/10.1103/PhysRevLett.113.046802} {\bibfield  {journal}
  {\bibinfo  {journal} {Phys. Rev. Lett.}\ }\textbf {\bibinfo {volume} {113}},\
  \bibinfo {pages} {046802} (\bibinfo {year} {2014})}\BibitemShut {NoStop}%
\bibitem [{Supp()}]{Supp}%
  \BibitemOpen 
  \bibfield  {title} {\bibinfo {title} {See {{Supplemental Material}} at [url]
  for the details of similarity between Hamiltonians, two-band Berry phase, non-Hermitian Creutz ladder model
  and proof for the same winding number for different random configurations. 
  which includes {{Ref}}. [29]}}\href@noop {\ } {\ }\BibitemShut {NoStop}%
\bibitem [{\citenamefont {Vanderbilt}(2018)}]{vanderbilt2018book}%
  \BibitemOpen
  \bibfield  {author} {\bibinfo {author} {\bibfnamefont {D.}~\bibnamefont
  {Vanderbilt}},\ }\href@noop {} {\emph {\bibinfo {title} {Berry {{Phases}} in
  {{Electronic Structure Theory}}: {{Electric Polarization}}, {{Orbital
  Magnetization}} and {{Topological Insulators}}}}}\ (\bibinfo  {publisher}
  {{Cambridge University Press}},\ \bibinfo {address} {{Cambridge}},\ \bibinfo
  {year} {2018})\BibitemShut {NoStop}%
\end{thebibliography}
%

\end{document}